# Utilisation des méthodes de Lee-Carter et Log-Poisson pour l'ajustement de tables de mortalité dans le cas de petits échantillons

**Frédéric PLANCHET*** **Vincent LELIEUR**[α]

**ISFA – Université Lyon 1** [β]
**WINTER & Associés** [γ]

RESUME

L’objectif de ce travail est d’étudier la mise en place de tables de mortalité prospectives à partir de faibles effectifs soumis au risque. Les modèles présentés sont les méthodes de Lee-Carter et log-Poisson respectivement. La faible taille des effectifs étudiés, particulièrement observée aux âges avancés, implique l’utilisation d’une méthode d’extrapolation des taux de mortalité aux âges élevés. Les méthodes Lee-Carter et log-Poisson constituent des modèles bi-dimensionnels prenant en compte à la fois l’année et l’âge pour déterminer les taux de mortalité. Les méthodes proposées sont appliquées à un portefeuille réel. Les tables prospectives ainsi construites permettent de projeter l’évolution des taux dans le futur, par extrapolation de la composante temporelle, et de comparer cette projection avec celle prévue pour la population française dans son ensemble. On détermine la meilleure méthode par la proximité des taux lissés par rapport aux taux bruts et surtout par la prudence de ces modèles pour le calcul de rentes viagères. Les résultats issus de ces méthodes sont également confrontés aux taux de mortalité obtenus par un ajustement logistique.

MOTS-CLEFS : Tables prospectives, extrapolation, lissage, rentes viagères, mortalité.

ABSTRACT

The aim of this paper is to study the construction of prospective mortality tables from a low number of persons subjected to risk. The presented models are the Lee-Carter and log-Poisson methods respectively. The low number of people subjected to risk, particularly noticed for the persons who are getting on, implies the use of an extrapolation method for the mortality rates. The Lee-Carter and log-Poisson methods constitute two-dimensional models, taking the year and the age into account to calculate the mortality rates. The methods suggested are applied to a real data set. The prospective tables, built in this way, allow to project the rates’ evolution in the future, extrapolating the temporal constituent. And then, it allows to compare this projection with the evolution predicted for the French population in its entirety. You determine the best method through the nearness of the smoothed rates in comparison with the raw rates and essentially through the caution of these models for the life annuities’ calculation. The results stemed from these methods are too confronted with the mortality rates obtained through a method of logistic fits.

KEYWORDS : Prospective tables, extrapolation, adjustment, life annuities, mortality.

---

* Frédéric Planchet est professeur associé de Finance et d’Assurance à l’ISFA (Université Lyon 1 – France) et actuaire associé chez WINTER & Associés. Contact : fplanchet@winter-associes.fr.

[α] Vincent Lelieur est actuaire à la CNAV. Contact : vincent.lelieur@cnav.fr.

[β] Institut de Science Financière et d’Assurances (ISFA) - 50, avenue Tony Garnier 69366 Lyon Cedex 07.

[γ] Winter & Associés – 43-47 avenue de la Grande Armée 75016 Paris et 18 avenue Félix Faure 69007 Lyon.

# Introduction

En France, la liquidation de capitaux sous forme de rentes n’a qu’un faible attrait sur les ménages. Pourtant, les rentes viagères constituent un mode de financement intéressant de la période de retraite, car elles garantissent contre le risque de longévité (*cf.* PANSARD [2005]). La mesure des engagements associés à ce type de prestation dépend fortement de l’évolution future de la mortalité (on rappelle que sur les trente dernières années en France on peut considérer que l’espérance de vie à la naissance a augmenté d’un trimestre par an en moyenne). La nécessité d’utiliser des tables de mortalité prospectives intégrant ce phénomène de dérive a été prise en compte par le législateur et des tables « de génération » ont été homologuées il y a une dizaine d’années. Ces tables, obtenues sur la base de la mortalité de la population féminine sur la période 1961-1987, servent depuis le 1er juillet 1993 à la tarification et au provisionnement de contrats de rentes viagères (immédiates ou différées). Elles sont en cours d’actualisation et des nouvelles tables devraient voir le jour en 2006.

Toutefois, en utilisant des tables exogènes au groupe considéré on se trouve confronté au risque de sous-provisionnement des rentes viagères (*cf.* BROUHNS et DENUIT [2001]) pour un groupe dont la mortalité serait inférieure à celle de la population de référence.

Lorsque la taille du groupe le permet, on peut donc vouloir s’orienter vers la construction de tables d’expérience spécifiques, avec la volonté de cerner ainsi tout comportement de la population sous risque qui serait significativement différent des tables réglementaires, ou, plus généralement, de références nationales standards. Dans une problématique de rentes, il est indispensable d’intégrer au modèle la dérive temporelle de la mortalité, ce qui rend la construction de tables d’expériences plus complexe.

Deux approches sont envisageables, qui utilisent les données des années antérieures sur la mortalité du groupe pour construire un ajustement des taux de décès bruts passés, puis des projections de la mortalité future :

- ✓ *Les modèles de construction « intrinsèque » (ou endogène)* : la démarche consiste ici à exploiter l’information contenue dans les taux bruts de mortalité pour ajuster un modèle représentant correctement les données et permettant une projection « réaliste ». Les effectifs relativement restreints à disposition peuvent toutefois nuire à l’identification de la dérive de mortalité dans une proportion délicate à quantifier (sur ces effets dans des modèles classiques de tables prospectives on pourra consulter LELIEUR [2005]).

- ✓ *Les modèles utilisant une référence externe de mortalité (ou exogène)* : une solution pour pallier les difficultés associées à des échantillons de taille réduite est de positionner la mortalité du groupe considéré par rapport à une référence externe, par exemple au moyen d’une régression logistique. Disposant d’un ensemble de tables de moments INSEE (féminine et masculine) historiques et prospectives l’idée est alors de « positionner » les tables du moment d’expérience par rapport à cet ensemble de tables.

Notre attention se porte dans ce travail pour l’essentiel sur la première classe de modèles, même si une alternative relevant de la seconde approche est présentée.

Les méthodes de construction intrinsèque récentes reposent sur des modèles bi-dimensionnels, qui « capturent » conjointement les deux composantes - l'âge et l'année - pour l'ajustement des taux de mortalité sur l'ensemble des années considérées. Il s'agit notamment du modèle de Lee-Carter (voir notamment LEE et CARTER [1992], LEE [2000], SITHOLE et al. [2000]) et de ses variantes de type log-Poisson (BROUHNS et al. [2002]). Les taux de mortalité pour les années futures se déduisent dans un second temps directement de la composante temporelle (paramétrique ou non paramétrique) du modèle prospectif retenu (On peut évidemment critiquer cette approche purement extrapolative ; on pourra par exemple consulter GUTTERMAN et VANDERHOOF [1999] sur ces questions).

Ces méthodes ont été initialement développées sur de très grands échantillons de population, à l'échelle d'un pays.

L'objectif du présent article est de présenter les ajustements qui doivent leur être apportés dans le cas de populations de taille plus réduite, typiquement à l'échelle d'un régime de retraite professionnel ou d'un portefeuille d'assurance. Le (relativement) faible effectif de la population soumise au risque conduit alors à des fluctuations d'échantillonnages importantes pour les taux bruts de mortalité. On met en évidence les problèmes sur les coefficients des paramétrisations par Lee-Carter (et log-Poisson) induits par ces fluctuations et on propose une méthode pour résoudre cette difficulté, en réduisant le nombre de paramètres du modèle.

Une autre conséquence du faible effectif soumis au risque de décès se manifeste aux âges les plus avancés. En effet, les effectifs décroissant rapidement avec l'âge au delà de 70 ans, les taux de mortalité aux âges les plus élevés sont très volatiles et aucune analyse statistique fiable ne peut en pratique être menée sur les données brutes au-delà de l'âge de 90 ans.

On se tourne alors vers des méthodes d'extrapolation des taux de mortalité aux âges les plus avancés. La méthode d'extrapolation de Coale-Kisker est la principale méthode d'extrapolation utilisée dans cette étude (COALE et KISKER [1990]).

Après avoir exposé l'approche par les modèles de Lee-Carter et log-Poisson, on effectue également un lissage de la mortalité brute en référence aux taux de mortalité INSEE, en effectuant un ajustement logistique. L'extrapolation est ensuite immédiate en utilisant les coefficients de la régression sur les tables prospectives exogènes. Ce modèle a été proposé dans SERANT [2005]. Il permet d'utiliser la structure apportée par une référence externe telle que l'INSEE et de positionner les taux d'expérience par rapport à cette référence.

Le choix du « meilleur » modèle s'appuie sur le test statistique du Chi-Deux, la proximité des taux ajustés avec ceux issus des données brutes et la prudence du coût du provisionnement des rentes viagères qu'il implique. Pour ne pas surcharger le présent article les applications au provisionnement des rentes n'y sont pas abordées. Le lecteur intéressé par ces questions pourra consulter LELIEUR [2005].

Les applications numériques sont issues de données sur la population masculine néo-calédonienne fournies pas l'INSEE.

Le présent travail est inspiré d'un travail de recherche effectué dans le cadre de l'Institut de Science Financière et d'Assurances de l'Université Lyon 1 par LELIEUR [2005].

# 1. La construction de tables prospectives

L'objectif de tables prospectives est de tenir compte des évolutions à venir de la mortalité ; les méthodes utilisées ici cherchent tout d'abord à ajuster les tendances passées, puis à les extrapoler à l'avenir. L'approche prospective consistant à intégrer dans l'avenir l'effet de progrès médicaux futurs n'est pas examinée ici.

On se propose d'ajuster les taux bruts à un modèle paramétrique, permettant d'une part de lisser les fluctuations d'échantillonnage et d'autre part de projeter l'évolution des taux dans le futur, par extrapolation. On dispose de taux bruts indicés par l'âge $x$ et l'année calendaire $t$, $\hat{q}_{xt}$, qui ont l'allure suivante :

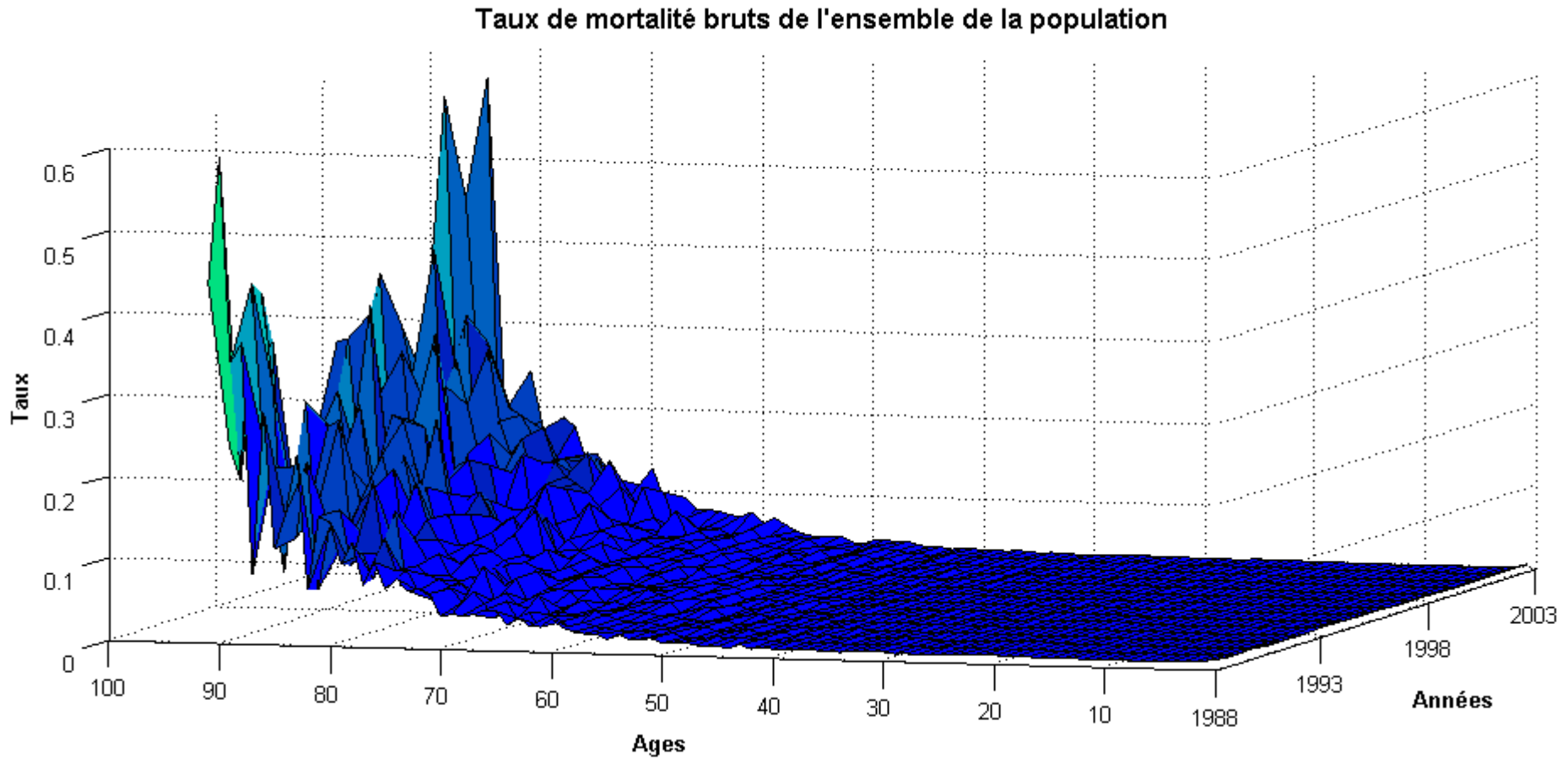

Fig. 1 : *Taux de décès bruts par année*

Le groupe étudié est constitué d'environ 100 000 personnes couvrant la plage d'âges 0-100 ans et observées pendant cinq ans. On dispose donc au maximum d'un millier de personne pour chaque âge observé, ce qui est peu. Les applications numériques présentées ici sont destinées à mettre en évidence graphiquement les phénomènes observés ; les valeurs numériques précises ne sont donc pas intégrée à ce papier, on a recours à un ensemble de graphiques.

Le passage du quotient de mortalité brut au taux instantané de mortalité, qui est la variable modélisée ici, s'effectue via une hypothèse sur la répartition des décès dans l'année (voir PLANCHET et THEROND [2006]) ; dans le cas où l'on fait l'hypothèse de constance du taux instantané dans chaque carré du diagramme de Lexis, on obtient l'estimateur suivant :

$$\mu^*_{xt} = -\ln\left(1-\hat{q}_{xt}\right). \qquad (1)$$

Avant d'utiliser des modèles d'ajustement, il convient de « fermer » les tables, les taux bruts au delà de 90 ans étant inexploitables. Différentes méthodes existent sur ce point (voir par exemple DENUIT et QUASHIE [2005] pour des développements sur ce sujet), et nous retiendrons ici la méthode de Coale et Kisker (COALE et KISKER [1990]) qui conduit à des résultats satisfaisants. Cette méthode consiste à extrapoler les taux instantanés de mortalité aux grands âges (jusqu'à $x$ = 110 ans par exemple) en se basant sur la formule[1] :

$$\hat{\mu}_x = \hat{\mu}_{65}.e^{g_x(x-65)}. \quad (2)$$

$g_x$ désignant le taux moyen de croissance de $\mu_x$ entre 65 et x ans. On calcule ainsi les coefficients $g_x$ jusqu'à un certain âge, puis on les extrapole afin de pouvoir recomposer les taux $\mu_x$. Coale et Kisker ont en effet remarqué empiriquement que les courbes des $g_x$ possèdent en général un pic aux alentours de 80 ans avant de décroître linéairement. Ils ont par conséquent proposé l'équation :

$$g_x = g_{80} + s(x-80),\ x \geq 80. \quad (3)$$

Finalement, on peut utiliser la formule suivante pour extrapoler au-delà de 80 ans les taux instantanés de mortalité :

$$\hat{\mu}_x = \hat{\mu}_{x-1}.e^{g_{80}+s(x-80)}, x \geq 80. \quad (4)$$

On utilise les valeurs de paramètres $s = -\frac{\ln(\hat{\mu}_{79} + 31.g_{80})}{465}$ et $g_{80} = \frac{\ln(\frac{\hat{\mu}_{80}}{\hat{\mu}_{65}})}{15}$.

## 1.1. La méthode de Lee-Carter

### 1.1.1. Présentation du modèle

Il s'agit d'une méthode d'extrapolation des tendances passées initialement utilisée sur des données américaines, qui est devenue rapidement un standard (voir l'article original LEE et CARTER [1992]). La modélisation retenue pour le taux instantané de mortalité est la suivante :

$$\ln \mu_{xt} = \alpha_x + \beta_x k_t + \varepsilon_{xt}, \quad (5)$$

avec les variables aléatoires $\varepsilon_{xt}$ i.i.d. selon une loi $N(0,\sigma^2)$ ; l'idée du modèle est donc d'ajuster à la série (doublement indicée par $x$ et $t$) des logarithmes des taux instantanés de décès une structure paramétrique (déterministe) à laquelle s'ajoute un phénomène aléatoire ; le critère d'optimisation retenu va consister à maximiser la variance expliquée par le modèle, ce qui revient à minimiser la variance des erreurs.

Le paramètre $\alpha_x$ s'interprète comme la valeur moyenne des $\ln(\mu_{xt})$ au cours du temps. On vérifie que $\frac{d\ln(\mu_{xt})}{dt} = \beta_x \frac{dk_t}{dt}$ et on en déduit que le coefficient $\beta_x$ traduit la sensibilité de la mortalité instantanée à l'âge $x$ par rapport à l'évolution générale $k_t$, au sens où

[1] On omet ici l'indice $t$ pour alléger les notations.

$\frac{d\,\mathbf{ln}(\mu_{xt})}{dk_t} = \beta_x$. En particulier, le modèle de Lee-Carter suppose la constance au cours du temps de cette sensibilité. Cette contrainte du modèle peut apparaître relativement forte :

- Pour tout age $x$ les quotients des variations relatives des taux de mortalité à des dates différentes ne dépendent pas de l'age $x$. Si la variation relative du taux de mortalité à 50 ans en 2000 était 80 % de ce quelle était en 1990 ce coefficient de 80 % est retenu pour tous les ages ;
- Pour une même date $t$ les quotients des variations relatives des taux de mortalité à des âges différents ne dépendent pas de la date $t$. Si en 2000 la variation relative du taux de mortalité à 20 ans est 50 % de la variation relative du taux à 50 ans ce coefficient de 50 % s'appliquera à toute date future ou passée.

Enfin, on peut remarquer que la forme du modèle implique l'homoscédasticité des taux de mortalité, ce qui est manifestement faux en pratique. Cet inconvénient sera examiné plus en détails *infra*. Afin de rendre le modèle identifiable, il convient d'ajouter des contraintes sur les paramètres ; on retient en général les contraintes suivantes :

$$\sum_{x=x_m}^{x_M} \beta_x = 1 \text{ et } \sum_{t=t_m}^{t_M} k_t = 0. \qquad (6)$$

On obtient alors les paramètres par un critère de moindres carrés (non linéaire) :

$$\left(\hat{\alpha}_x, \hat{\beta}_x, \hat{k}_t\right) = \mathbf{arg\,min} \sum_{x,t} \left(\mathbf{ln}\, \mu_{xt}^{*} - \alpha_x - \beta_x k_t\right)^2 \qquad (7)$$

Il convient donc de résoudre ce programme d'optimisation, sous les contraintes d'identifiabilité. Le nombre de paramètres à estimer est élevé, il est égal à $2 \times (x_M - x_m + 1) + t_M - t_m - 1$. Les algorithmes de résolution ne sont pas repris ici, le lecteur intéressé pourra consulter les nombreux articles décrivant ce modèle, par exemple BROUHNS et DENUIT [2001]. Une présentation détaillée est également proposée dans PLANCHET et THEROND [2006].

### 1.1.2. Application numérique

Avec les données présentées *supra* (Fig. 1 : ci-dessus), le modèle de Lee-Carter conduit à la surface suivante :

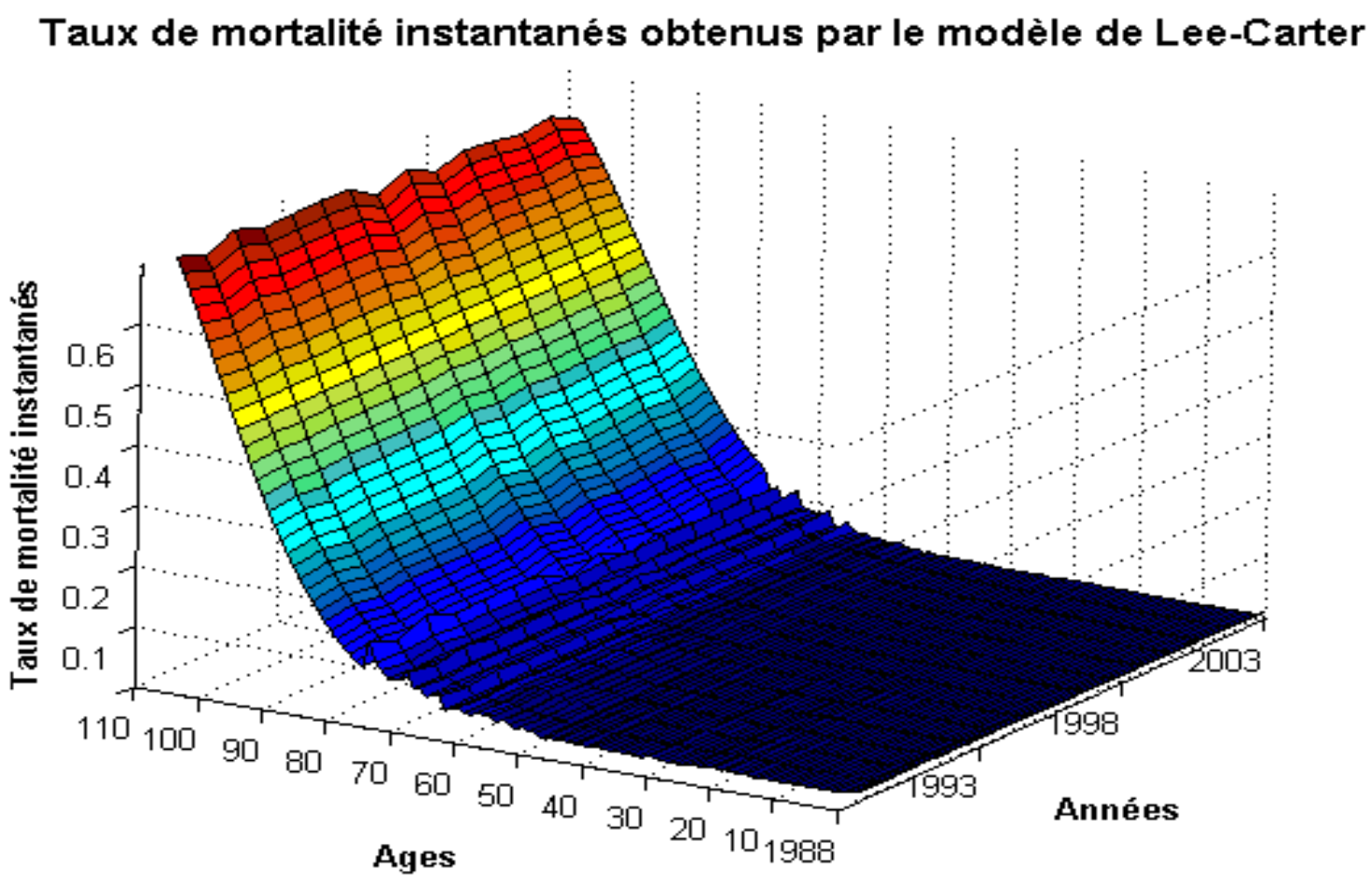

Fig. 2 : *Taux instantanés de mortalité par année, par le modèle de Lee-Carter.*

Les taux de mortalité instantanés obtenus par le lissage de Lee-Carter sont très volatiles. La volatilité des taux entre les années, et entre les âges, est particulièrement visible à partir de 65 ans. Ces fortes variations montrent que le lissage des taux de mortalité par le modèle de Lee-Carter n'est pas très efficace, lorsqu'on laisse les paramètres du modèle libres.

On peut observer que cette irrégularité est aussi accentuée par le fait qu'après une première estimation des paramètres par le critère des moindres carrés, le modèle propose de recaler les nombres de décès prédits par le modèle sur les nombres de décès observés ; si cette contrainte apparaît relativement naturelle sur de très grand groupes pour lesquels on peut estimer que les fluctuations d'échantillonnages sont négligeables, elle se justifie moins sur un groupe de taille plus réduite, dans lequel les taux bruts sont encore « bruités ».

Ce modèle présente par ailleurs un certain nombre de limitations, décrites ci-après.

### 1.1.3. Les limites du modèle de Lee-Carter

Le modèle de Lee-Carter repose sur l'hypothèse d'homoscédasticité des taux de mortalité, ce qui constitue une hypothèse forte et peu réaliste : en effet, la variance des taux de décès croît aux âges élevés, du fait notamment de la baisse des effectifs de survivants. On peut illustrer ce fait de deux manières ; tout d'abord, on considère la population française au 01/01/2005, que l'on suppose mourir selon la table TV 1999/2001. La variance des taux de décès bruts que l'on observerait peut être approchée par $\frac{q_x(1-q_x)}{L_x}$, et on constate l'évolution suivante :

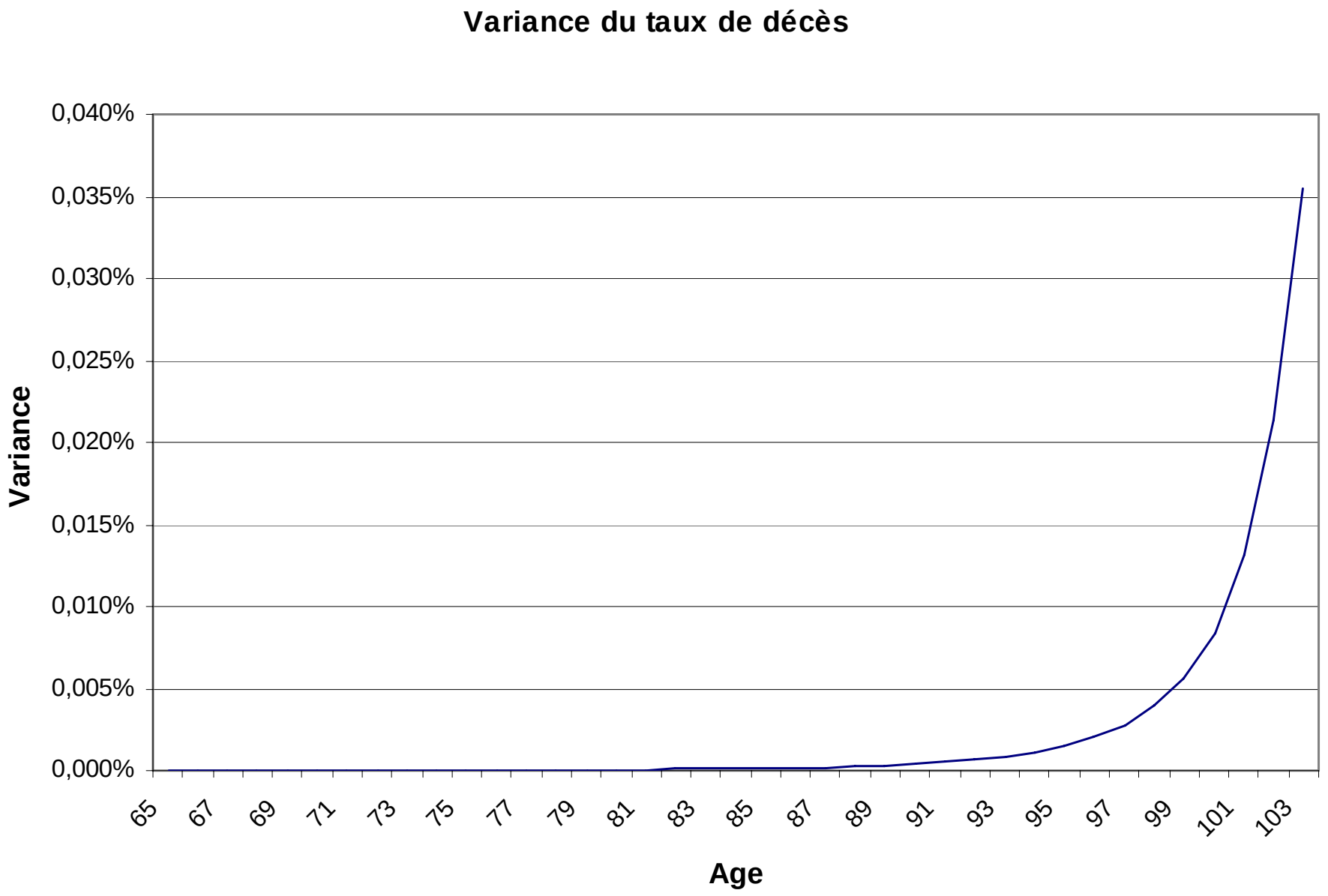

Fig. 3 : *Variance du taux de décès avec l'âge*

On note une très forte augmentation après 85 ans. De manière plus directe, lorsque l'on effectue un ajustement par la méthode de Lee-Carter, on peut analyser la variance des résidus, et confronter les observations à l'hypothèse d'hétéroscédasticité. On obtient des graphiques à l'allure suivante[2] :

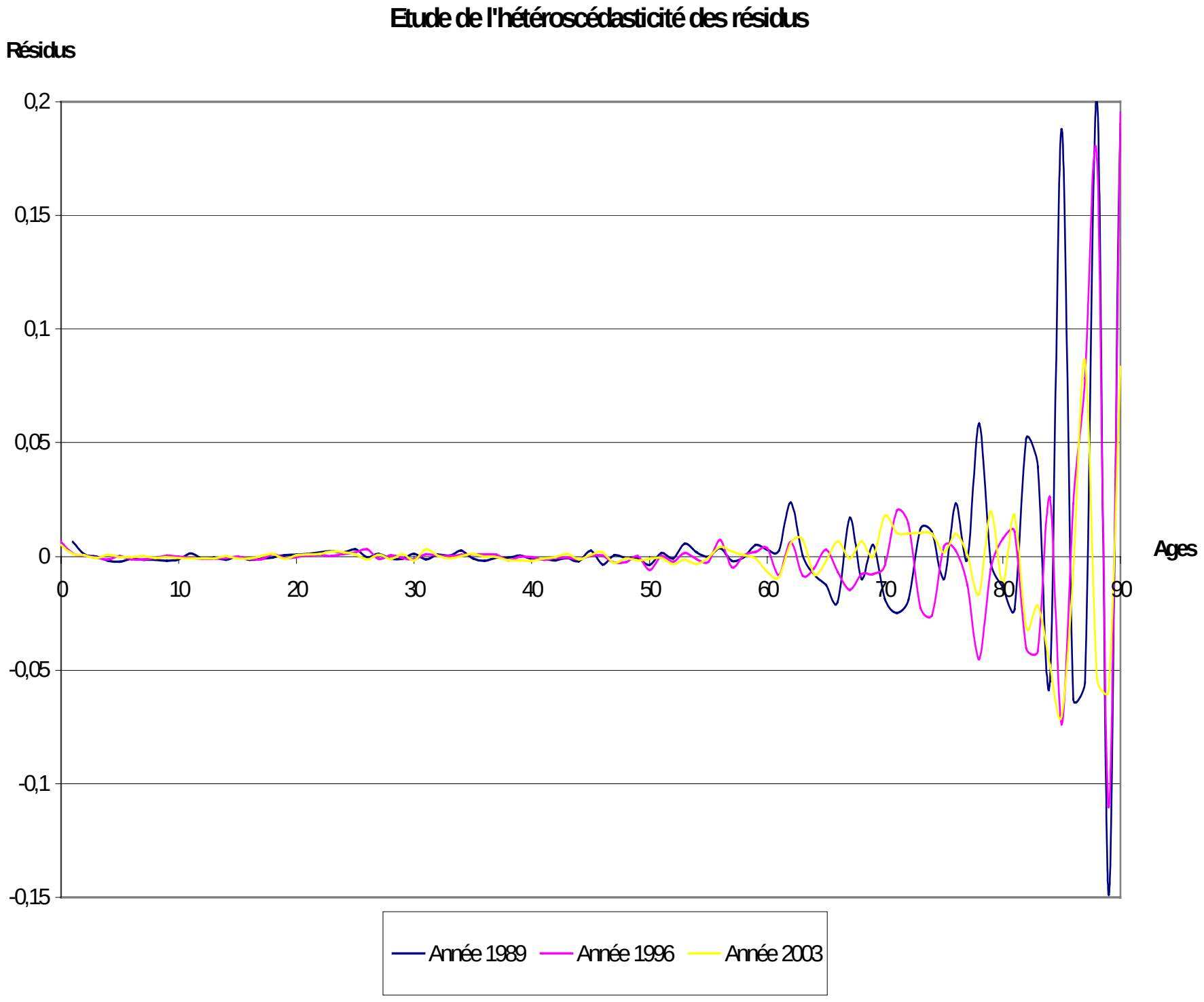

Fig. 4 : *Analyse des résidus dans le modèle de Lee-Carter*

[2] Voir LELIEUR [2005].

On peut par ailleurs noter que le critère de sélection des paramètres optimaux dans le modèle de Lee-Carter n'a pas de justification probabiliste, et chercher alors à modifier l'approche pour obtenir un critère de détermination des paramètres de type « maximum de vraisemblance », qui permettra de bénéficier des « bonnes propriétés » de cette classe d'estimateurs (convergence, efficacité asymptotique, normalité asymptotique, etc.).

Tout ceci incite à rechercher un autre modèle palliant ces inconvénients. Le modèle log-Poisson, proposé par BROUHNS et al. [2002], est une adaptation du modèle de Lee-Carter précisément destinée à pallier ces deux limites.

## 1.2. Le modèle log-Poisson

### 1.2.1. Présentation du modèle

L'idée est de modéliser le nombre de décès à l'âge $x$ l'année $t$, $D_{xt}$, par une loi de Poisson, en supposant que $D_{xt}$ suit une loi de Poisson de paramètre $L_{xt}\mu_{xt}$ avec $\mu_{xt} = \exp(\alpha_x + \beta_x k_t)$. L'expression du taux de décès instantané est identique à celle proposée dans le modèle de Lee-Carter, avec la même interprétation des différents paramètres. En particulier, le modèle ne sera identifiable qu'avec des contraintes sur les paramètres, et on peut retenir les mêmes que celles utilisées par Lee et Carter. Enfin, on peut noter que passer du modèle de Lee-Carter à ce modèle poissonien revient à passer d'un modèle linéaire à un modèle linéaire généralisé avec le logarithme comme fonction de lien[3].

Comme on a $P(D_{xt} = d) = \frac{(L_{xt}\mu_{xt})^d}{d!} \exp(-L_{xt}\mu_{xt})$ avec $\mu_{xt} = \exp(\alpha_x + \beta_x k_t)$, la log-vraisemblance[4] du modèle s'écrit (à une constante additive près) :

$$\ln L(\alpha, \beta, k) = \sum_{x,t} \left\{ D_{xt}(\alpha_x + \beta_x k_t) - L_{xt} \exp(\alpha_x + \beta_x k_t) \right\} \qquad (8)$$

On dispose donc d'une expression simple de la log-vraisemblance ; les équations de vraisemblance n'ont pas de solution analytique du fait de la présence du terme non linéaire $\beta_x k_t$ et doivent être résolues numériquement ; on peut par exemple utiliser un algorithme de Newton-Raphson et utiliser comme fonction objectif $F$ à annuler le vecteur des scores $\left( \frac{\partial L}{\partial \alpha}, \frac{\partial L}{\partial \beta}, \frac{\partial L}{\partial k} \right)'$. On notera que le nombre de paramètres du modèle est identique au cas du modèle précédent, et que leur interprétation est identique.

---

[3] On pourra se reporter à RENSHAW [1991]. Cette remarque peut conduire à d'autres modèles en jouant sur la distribution de référence du modèle linéaire généralisé.

[4] Il ne s'agit d'une vraisemblance que si on utilise les effectifs sous risque réels ; si on normalise les effectifs en partant d'un effectif initial de $L_0$, on obtient une pseudo-vraisemblance.

### 1.2.2. Application numérique

Avec les données de notre exemple, et en travaillant sur les effectifs sous risque réel (et donc en situation d'estimation par le maximum de vraisemblance), le modèle log-Poisson conduit à :

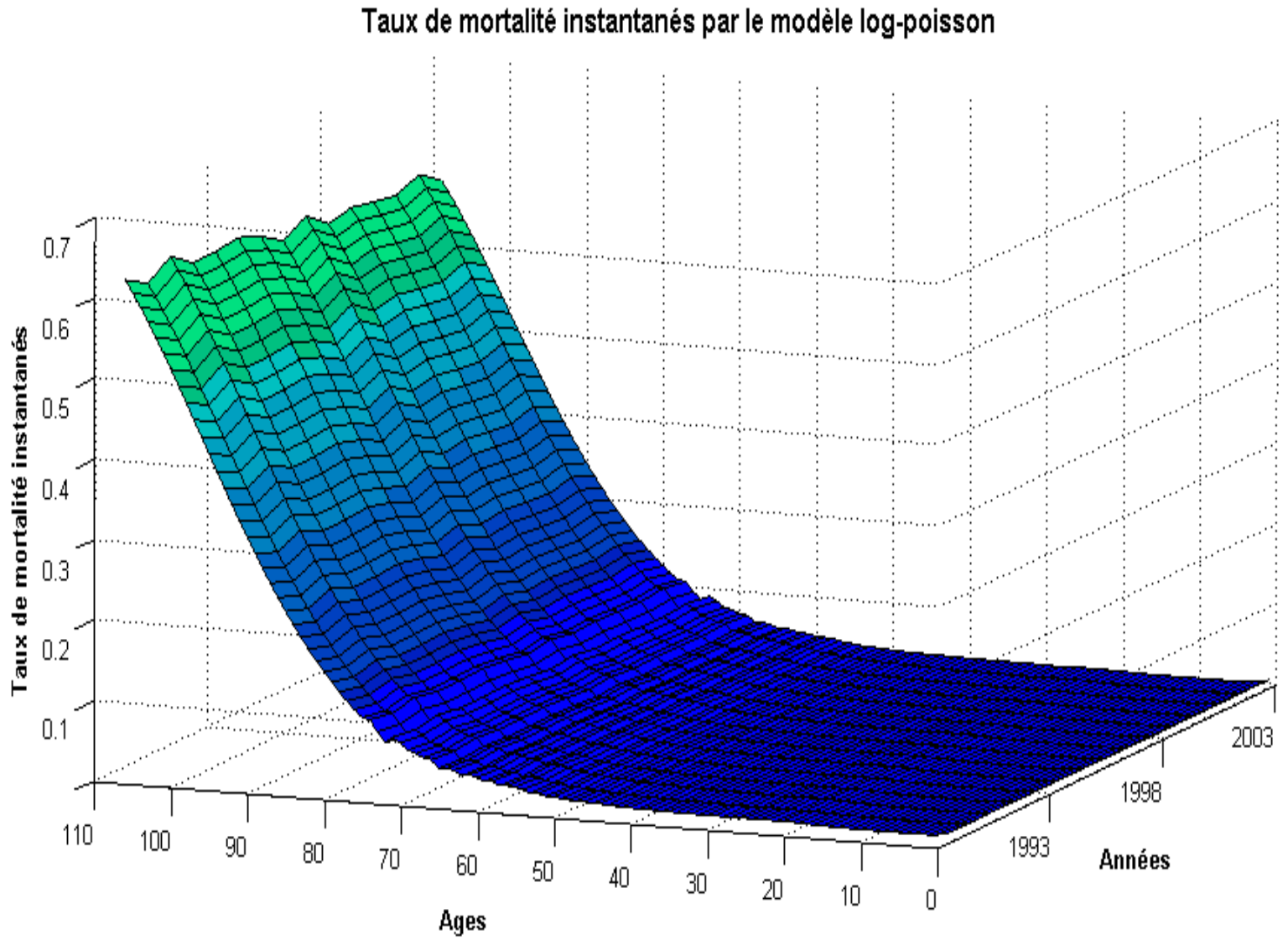

Fig. 5 : *Taux de mortalité lissés par le modèle log-poisson .*

On a observe les améliorations apportées par la régression poissonienne par rapport au modèle de Lee-Carter (en termes de régularité des taux et des paramètres). Cependant, il subsiste une certaine volatilité des paramètres affectant la régularité des taux calculés. Il importe donc de corriger ce défaut.

L'allure des courbes sur les paramètres $\alpha$, $\beta$ et $k$ met en évidence une variabilité des coefficients, celle-ci entraînant une instabilité des taux de mortalité obtenus par le modèle. On choisit par conséquent de réduire le nombre de paramètres[5]. Cette remarque provient de l'allure de la courbe des $\alpha$ en fonction de l'âge. On propose la paramétrisation polynomiale :

$$\alpha_x = a_1 + b_1 x + c_1 x^2 + d_1 x^3 \quad (9)$$

Elle conduit au graphe suivant, sur lequel on rapproche les valeurs du paramètres $\alpha$ dans le modèle libre et dans le modèle contraint (on indique à titre d'information le coefficient $R^2$ associé) :

[5] Cette démarche peut être mise en œuvre sur le modèle de Lee-Carter, mais elle ne corrige pas les défauts de base de ce modèle, c'est pourquoi elle n'a pas été présentée ici.

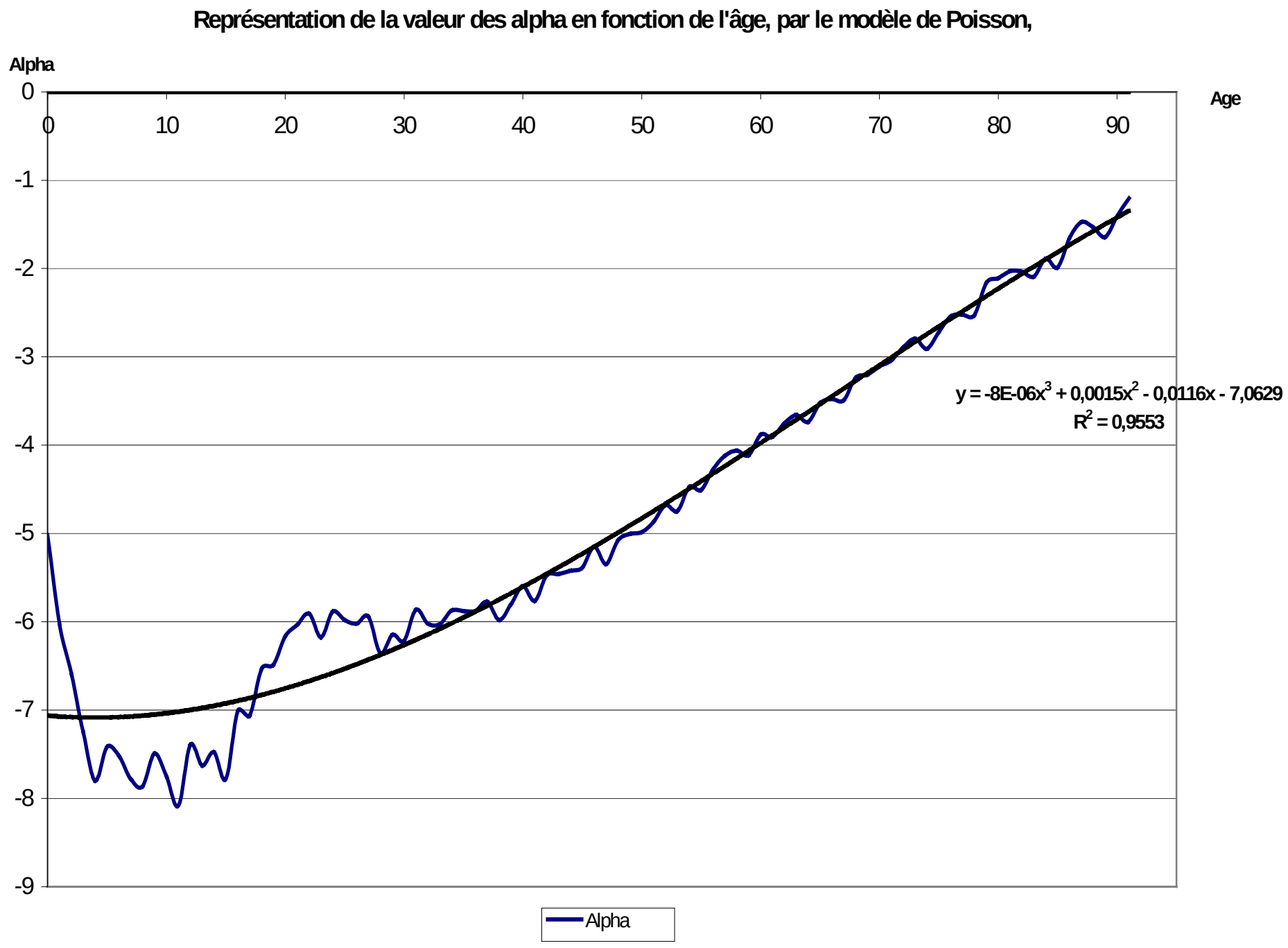

Fig. 6 : *Approximation polynomiale des paramètres alpha obtenus par les modèles log-poisson.*

La même démarche pour les coefficients $\beta$, contraint par un polynôme de degré 3 fournit les résultats suivants :

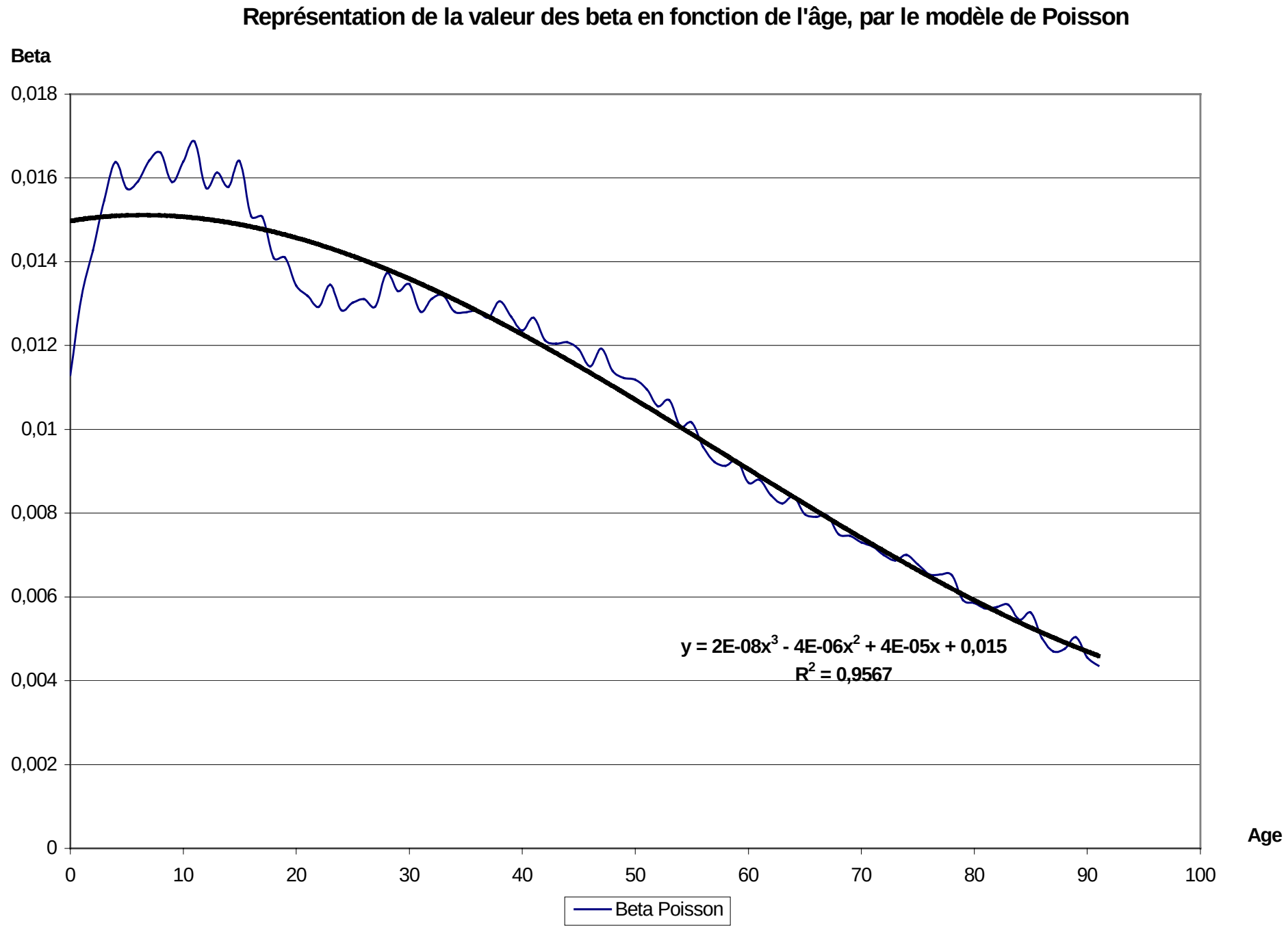

Fig. 7 : *Approximation polynomiale des paramètres béta obtenus par les modèles log-poisson.*

Le coefficient $k$ est simplement contraint à évoluer linéairement. On construit ainsi, par le maximum de vraisemblance, la surface (régulière par construction) suivante :

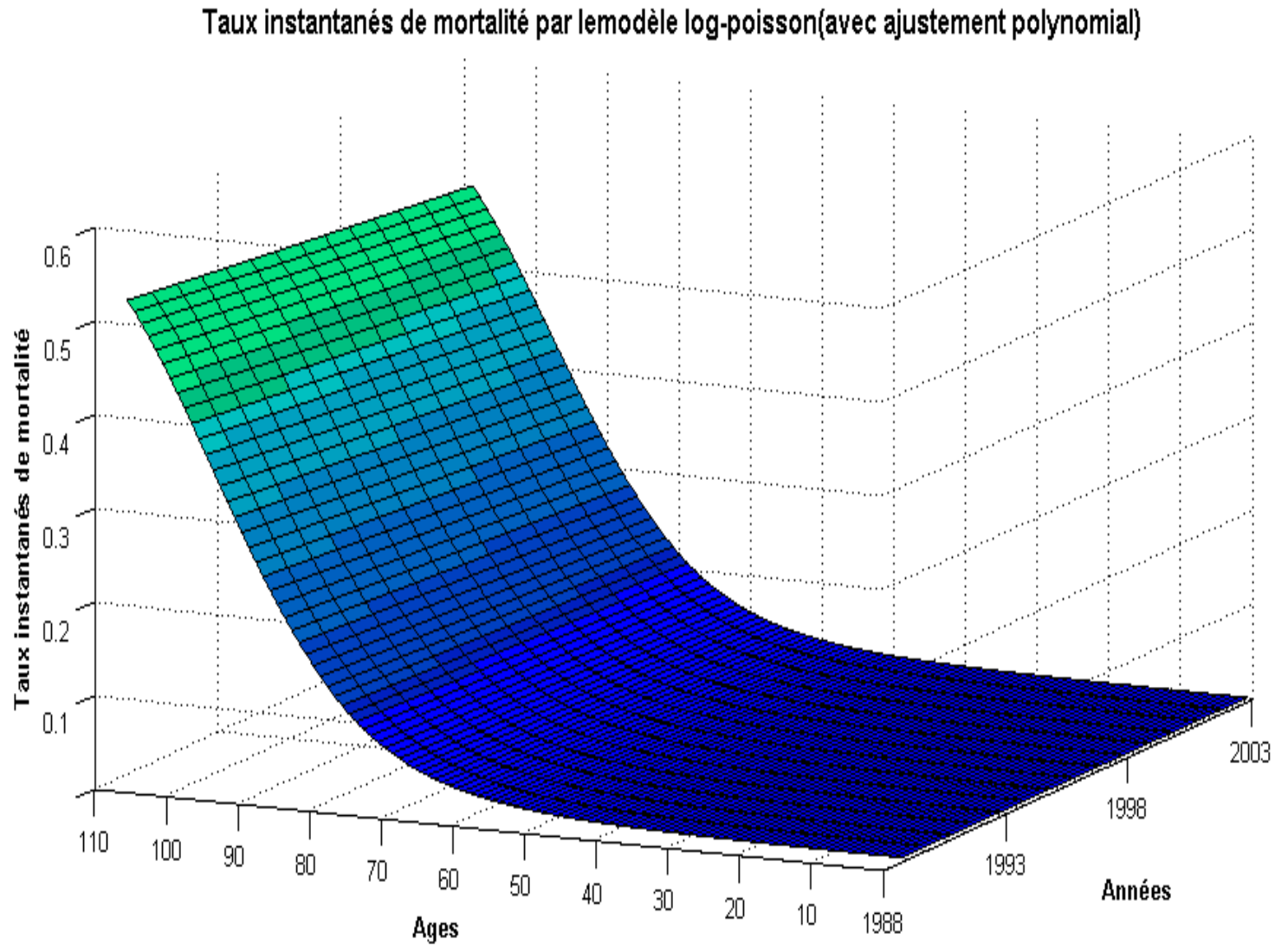

Fig. 8 : *Taux de mortalité instantanés par le modèle log-poisson (où les paramètres ont une structure polynomiale).*

Remarque : on peut en pratique retenir deux démarches qui conduisent sensiblement aux mêmes résultats ; la première, mise en œuvre ici, consiste à restreindre le nombre de paramètres à estimer et à les estimer par le maximum de vraisemblance. La seconde consiste à ajuster les moindres carrés des polynômes aux courbes des coefficients standards estimés par maximum de vraisemblance sur le modèle libre. La première solution est bien entendu plus satisfaisante d'un point de vue formel.

L'extrapolation des taux futurs est alors directe, *via* le prolongement de la tendance (linéaire) *k*. On notera incidemment que la poursuite d'une tendance linéaire sur le long terme peut apparaître irréaliste, mais on ne la remet pas en cause ici car elle est de nature à conduire à une majoration des engagements du régime, et est donc *a priori* une hypothèse prudente.

On obtient ainsi un modèle robuste, reposant sur une approche « maximum de vraisemblance » et avec un petit nombre de paramètres (6 = 4 + 4 + 2), ce quoi constitue une situation *a priori* favorable pour conduire la démarche prospective (voir sur ce sujet SERANT [2005] qui discute le lien entre paramétrisation et pouvoir prédictif d'un modèle). De plus l'approche par maximum de vraisemblance sur un nombre réduit de paramètre permet d'exploiter les propriétés asymptotiques de ces estimateurs (efficacité, convergence, tests de Chi-deux associés, etc.).

Enfin, on peut observer que l'on obtient ainsi une forme paramétrique simple de la fonction $\mu_{xt}$ en fonction de *x* et de *t*, ce qui conduit à des formules fermées pour le calculs des provisions à chaque âge (et à des calculs valables aux âges non entiers).

$$a_{xt} = \frac{1}{L_{xt}} \int_0^{+\infty} L_{xt}(h) \mathbf{exp}(-rh)\, dh, \qquad (10)$$

avec $r = \ln(1+i)$ la version continue du taux technique et $L_{xt}(h) = \exp\left(-\int_0^h \mu(x+\theta, t+\theta)\,d\theta\right)$ et $\mu_{xt} = \exp(\alpha_x + \beta_x k_t)$. Le calcul de cette intégrale ne pose pas de problème.

Ainsi, le modèle log-Poisson contraint présente un certain nombre de « bonnes propriétés » et améliore notablement, dans le cas des petits échantillons étudié ici, les résultats obtenus par rapport à l'application directe du modèle de Lee-Carter. Toutefois, l'extrapolation des taux de décès aux grands âges, qui définit la structure de la table après 80 ans, peut apparaître relativement arbitraire. L'utilisation d'un modèle « endogène » est ici délicate compte tenu de la faible quantité d'information apportée par les données.

Pour contourner cette difficulté, il peut alors apparaître naturel de chercher à s'appuyer sur une structure exogène existante ; l'INSEE fournit ainsi une série de table du moment pour l'ensemble du 20$^{e}$ siècle, avec une projection jusqu'en 2050. Cet ensemble de table va alors nous servir à positionner les données d'expérience, selon une logique semblable à ce que propose par exemple le modèle de Cox.

Ce modèle est présenté ci-après.

## 1.3. Un modèle basé sur une référence INSEE

### 1.3.1. Présentation du modèle

Dans le choix d'un modèle susceptible de structurer un jeu de données historiques, la « flexibilité » du modèle et par la même sa fidélité aux données est directement liée aux nombres de paramètres introduits. Le choix d'un modèle très flexible se fait le plus souvent au détriment des qualités prédictives de celui-ci (un modèle totalement non paramétrique n'autorise aucune prédiction).

Les modèles de Lee-Carter ou Log-Poisson standards peuvent de ce fait paraître très paramétrés. Au surplus, dans le contexte de données de portefeuilles, dont le volume est sensiblement inférieur à ce que l'on peut obtenir comme taille de population sous risque à l'échelle d'un pays, le nombre élevé de paramètres du modèle conduit, on l'a vu, à des irrégularités conséquences de fluctuations d'échantillonnage.

De plus, le modèle Log-Poisson, même contraint comme nous l'avons proposé *supra*, peut conduire à sous-évaluer notablement les taux de mortalité des ages élevés[6] (à partir de 85-90 ans). En effet, l'algorithme de référence construit sur une approche maximum de vraisemblance favorise les premiers âges (les plus « jeunes). Enfin, la relation $\mu_{xt} = -\ln(1-q_{xt})$ repose que l'hypothèse de constance du taux instantané de décès entre deux âges entiers, hypothèse discutable aux âges élevés.

[6] Ce défaut est toutefois à relativier dans le cas d'une application à des rentiers car il va dans le sens d'une estimation prudente de la mortalité future.

Dans ce contexte il peut être utile de se tourner vers des modèles alternatifs « naturellement » moins paramétrés et ne nécessitant pas l'hypothèse de constance du taux instantané sur chaque carré du diagramme de Lexis.

On choisit d'exprimer l'influence de l'âge $x$ et de l'année $t$ sur les taux de mortalité $q_x(t)$ au travers du « logit » des taux de décès :

$$\mathbf{lg}_x(t) = \mathbf{ln}\left(q_{xt} / (1 - q_{xt})\right). \qquad (11)$$

Le logit pour des taux de mortalités faibles est peu différent de la variable $\mathbf{ln}(\mu_{xt})$ du modèle de Lee-Carter (ou log-Poisson) mais il peut être sensiblement différent pour des âges élevés. Il présente l'avantage de varier dans $]-\infty, +\infty[$, ce qui simplifie la mise en œuvre de modèles de régression. La forme typique d'un logit est la suivante (obtenue avec la TV 1999/2001) :

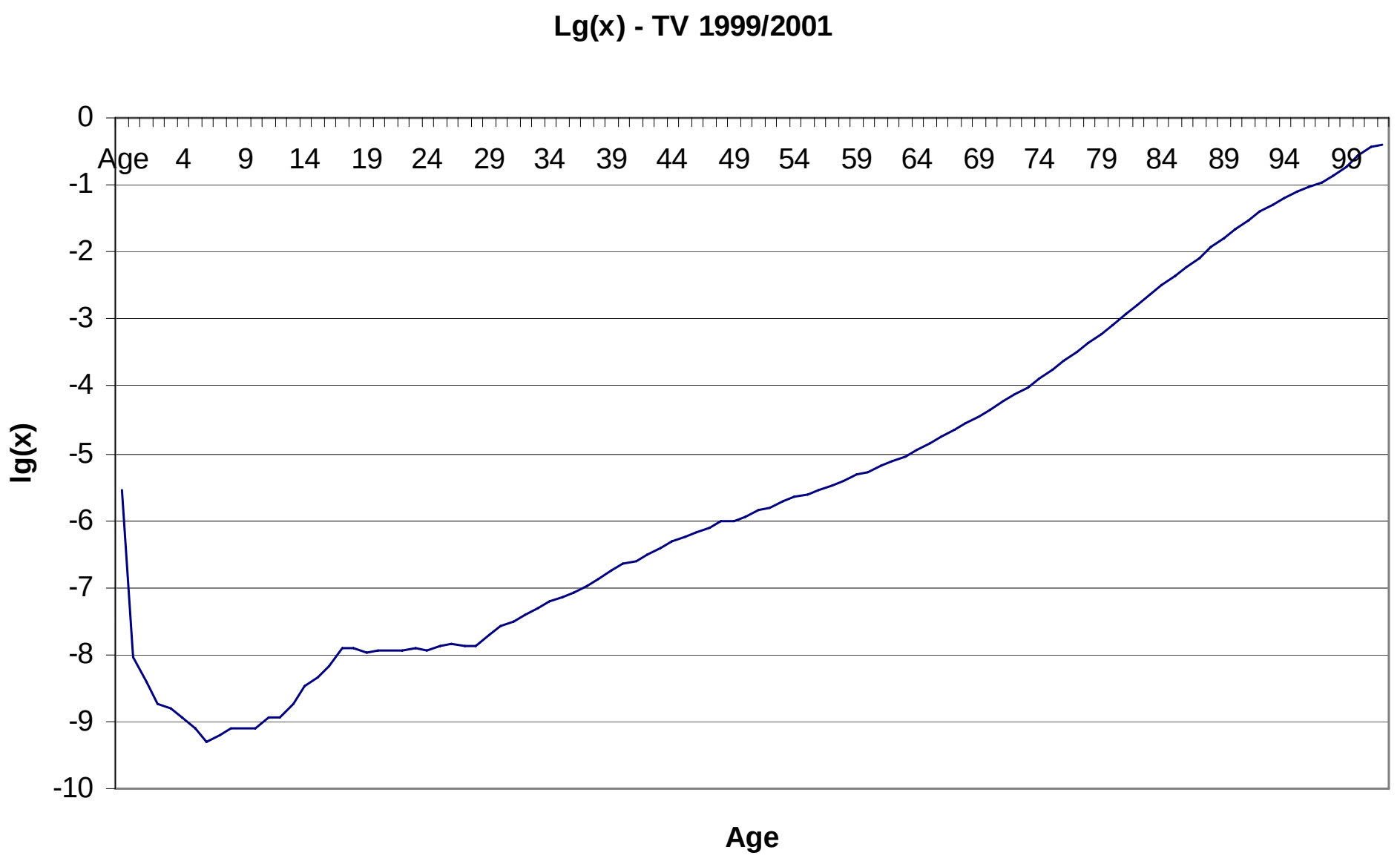

Fig. 9 : *Logit du taux de décès en fonction de l'âge*

Lorsque l'on souhaite positionner une table par rapport à une autre, il peut apparaître naturel d'effectuer la régression des logits des taux bruts sur les logits de la table de référence, ce qui conduit au modèle suivant :

$$\mathbf{lg}_x(t) = a_t \, \mathbf{lg}_x^{ref}(t) + b_t + \varepsilon_{xt}. \qquad (12)$$

La mise en œuvre de cette approche si l'on retient un critère de type « moindres carrés » est très simple, puisqu'il s'agit d'une régression linéaire dans le cadre d'un modèle linéaire ordinaire. On dispose donc d'une expression explicite des paramètres $a$ et $b$ (supposés dans l'étude indépendants de $t$). Elle permet au surplus une extrapolation aisée des logits des taux d'expérience dans les plages d'âge pour lesquelles les données d'expérience sont insuffisantes.

On a toutefois choisi ici d'adapter le critère d'optimisation utilisé pour tenir compte du contexte d'utilisation des tables en retenant plutôt :

$$(\hat{a},\hat{b}) = \mathbf{arg\,min}[e_{60}^{liss\acute{e}}(a,b) - e_{60}^{nonliss\acute{e}}]. \qquad (13)$$

sous la contrainte suivante :

$$e_{60}^{liss\acute{e}}(a,b) - e_{60}^{nonliss\acute{e}} > 0. \qquad (14)$$

où $e_{60}^{liss\acute{e}}(a,b)$ désigne l'espérance de vie résiduelle à 60 ans, fonction des paramètres $a$ et $b$, calculée à partir de la régression sur les logits ; $e_{60}^{nonliss\acute{e}}$ désigne l'espérance de vie résiduelle à 60 ans calculée à partir des données brutes. En d'autres termes, on cherche la paramétrisation conduisant aux provisions (à taux 0) à l'âge de 60 ans aussi proches que possible des provisions évaluées sur les taux bruts, en étant supérieures à ces dernières.

La résolution numérique de ce programme d'optimisation ne pose pas de problème particulier.

### 1.3.2. Application numérique

Le graphe suivant montre l'ensemble des taux de mortalité lissés obtenus pour les années 1989 à 2003, à l'aide de l'ajustement des taux logistiques INSEE.

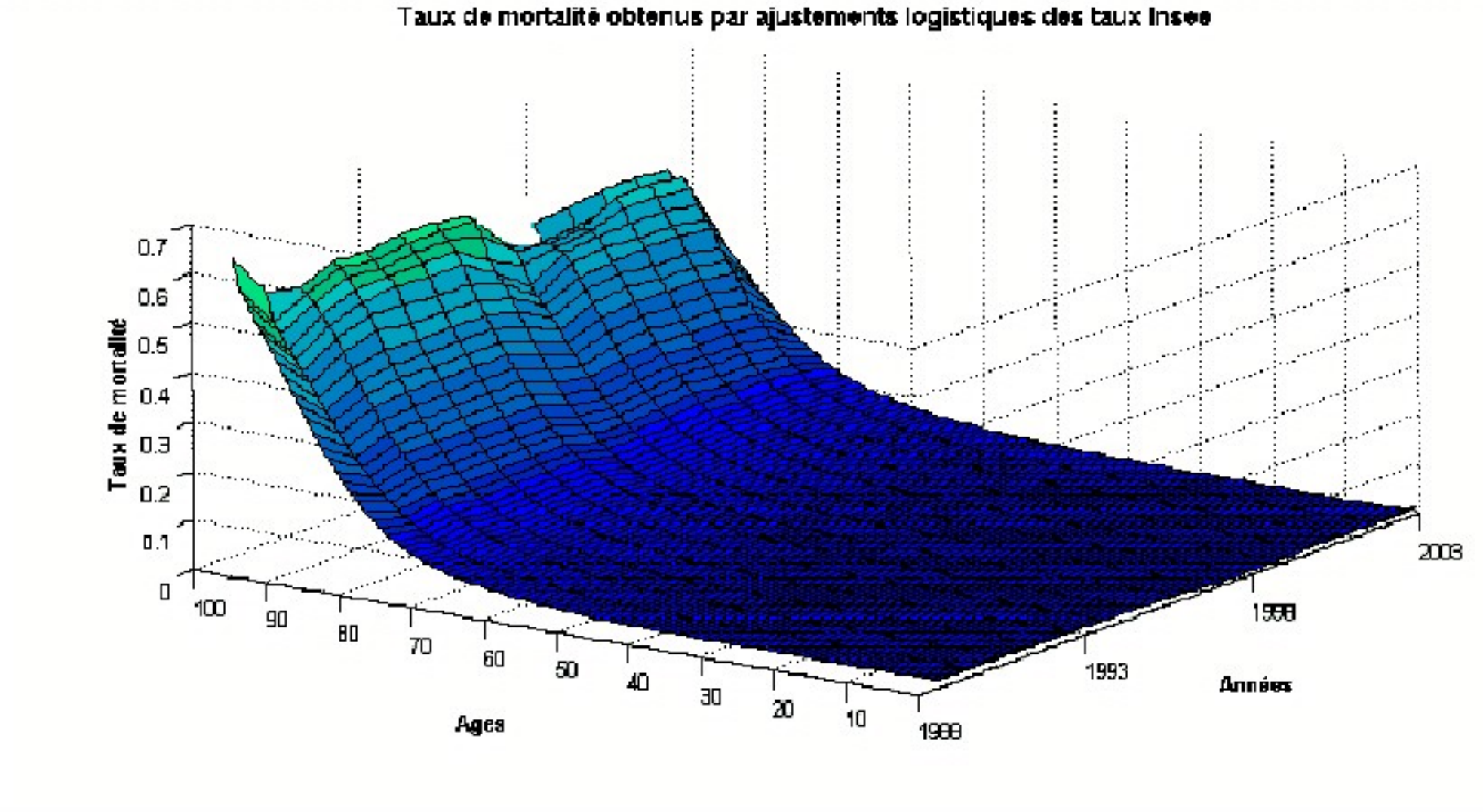

Fig. 10 : *Taux de mortalité obtenus par ajustement logistique (à partir des taux Insee).*

On constate qu'au-delà de 70 ans, les taux de mortalité lissés obtenus sont assez volatiles d'une année sur l'autre, contrairement aux résultats constatés avec les modèles de Lee-Carter et de Poisson, munis de l'ajustement polynomial des paramètres. Afin d'illustrer plus précisément la qualité du lissage obtenu à l'aide des taux logistiques INSEE, on présente ci-dessous les résultats (taux lissés) pour l'année 1989 :

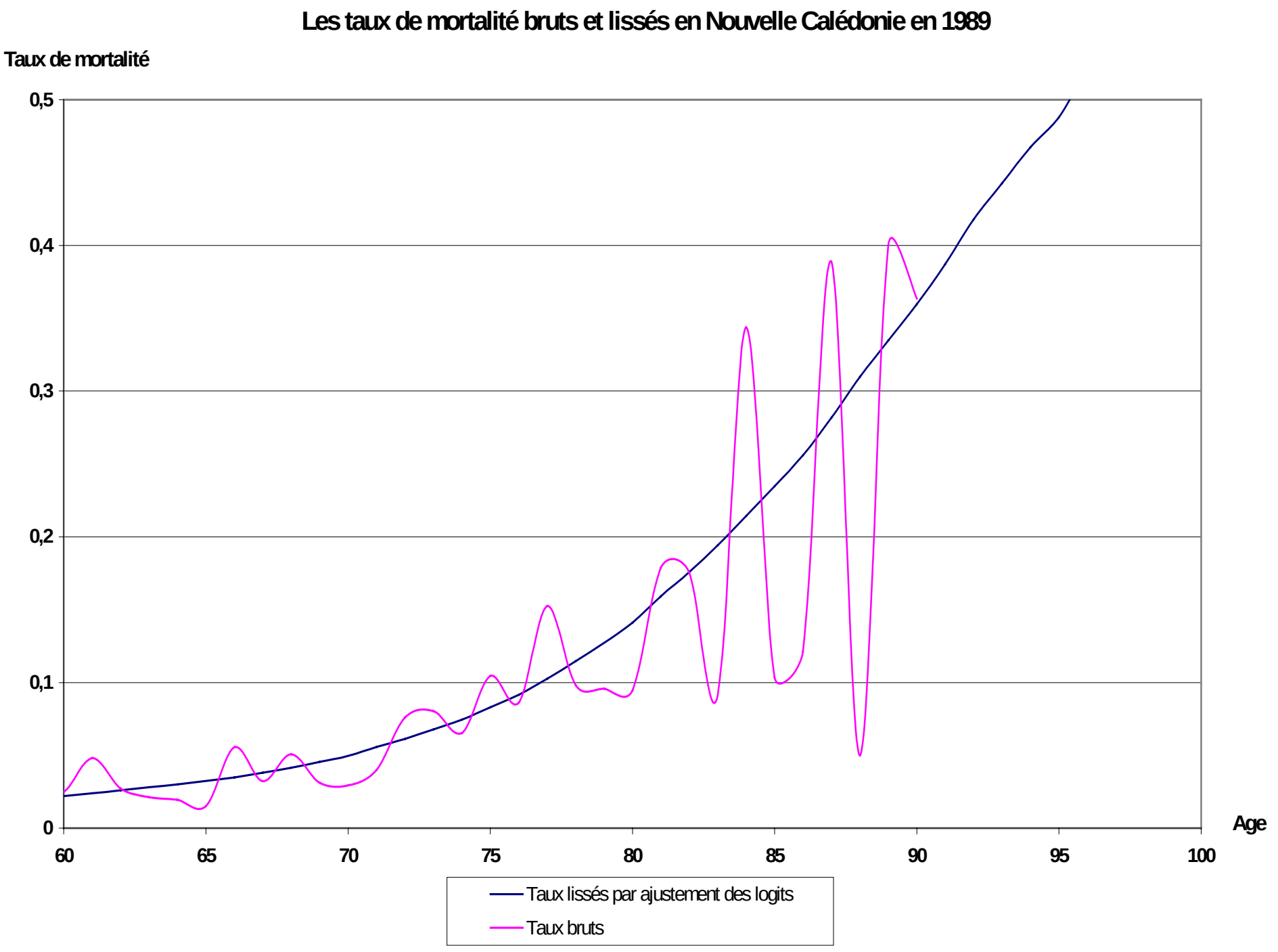

Fig. 11 : *Comparaison des taux de mortalité obtenus par ajustement logistique avec les taux de mortalité bruts, en 1989.*

On a présenté les résultats à partir de 60 ans, puisque c'est à partir de cet âge que sont calculés les coefficients de la régression. En 1989, le lissage par régression logistique est correct, les taux lissés semblent bien ajuster la mortalité observée.

# 2. Synthèse des résultats

Le choix de la collection de tables à retenir est guidé par la traditionnelle analyse des critères de fidélité aux données et de régularité, mais aussi, dans le cas présent d'un souci d'une approche prudente de l'évolution future du risque. Ces deux aspects sont évoqués ici.

## 2.1. Ajustement aux taux bruts

On calcule la déviation entre les décès observés (par les données brutes) et les *décès théoriques* (par le modèle de lissage employé, Lee-Carter ou Poisson) pour chaque génération (de 1989 à 2003) et chaque classe d'âge (de 0 à 89 ans[7]). Le calcul de la déviation (écart à l'indépendance entre les décès observés et théoriques) pour un âge $x$ et une année $t$ s'effectue par la formule suivante :

$$\chi^2_{x,t} = \frac{(D_{x,t} - D^*_{x,t})^2}{D^*_{x,t}}, \qquad (15)$$

où $D_{x,t}$ représente le nombre de décès observés (issus directement des taux bruts) à l'âge $x$ et pour l'année $t$; $D^*_{x,t}$ représentant les *décès théoriques* par application de la méthode de lissage de Lee-Carter ou de Poisson. On observera que cette statistique n'est *a priori* pas un Chi-deux, même asymptotiquement. Elle n'est utilisée ici que comme mesure de l'écart entre les taux ajustés et les taux bruts, et non comme une statistique de tests.

[7] On ne calcule pas la déviation entre les décès observés et théoriques au-delà de 89 ans, car les données brutes ont parfois des effectifs de tailles nulles pour les âges très avancés.

La valeur globale de la déviation entre les décès observés et théoriques est égale à la somme de tous les Chi-deux locaux:

$$\chi_t = \sum_x \chi^2_{x,t} = \sum_x \frac{(D_{x,t} - D^*_{x,t})^2}{D^*_{x,t}}, \qquad (16)$$

On indique à titre indicatif les statistiques du Chi-deux obtenues avec les différents modèles utilisés.

Les paramètres de la régression linéaire des logits néo-calédoniens par les logits INSEE ont été déterminés de sorte qu'il y ait égalité entre l'espérance de vie brute et lissée à 60 ans. De plus, comme ces tables sont destinées à être utilisées dans le cadre du provisionnement de rentes viagères, on décide de restreindre l'étude du Chi-deux sur les âges avancés, à partir de 60 ans. On partage la plage d'études [60 ans, 89 ans] en six classes comprenant chacune cinq années. Les résultats obtenus sont les suivants, pour chaque année d'observation :

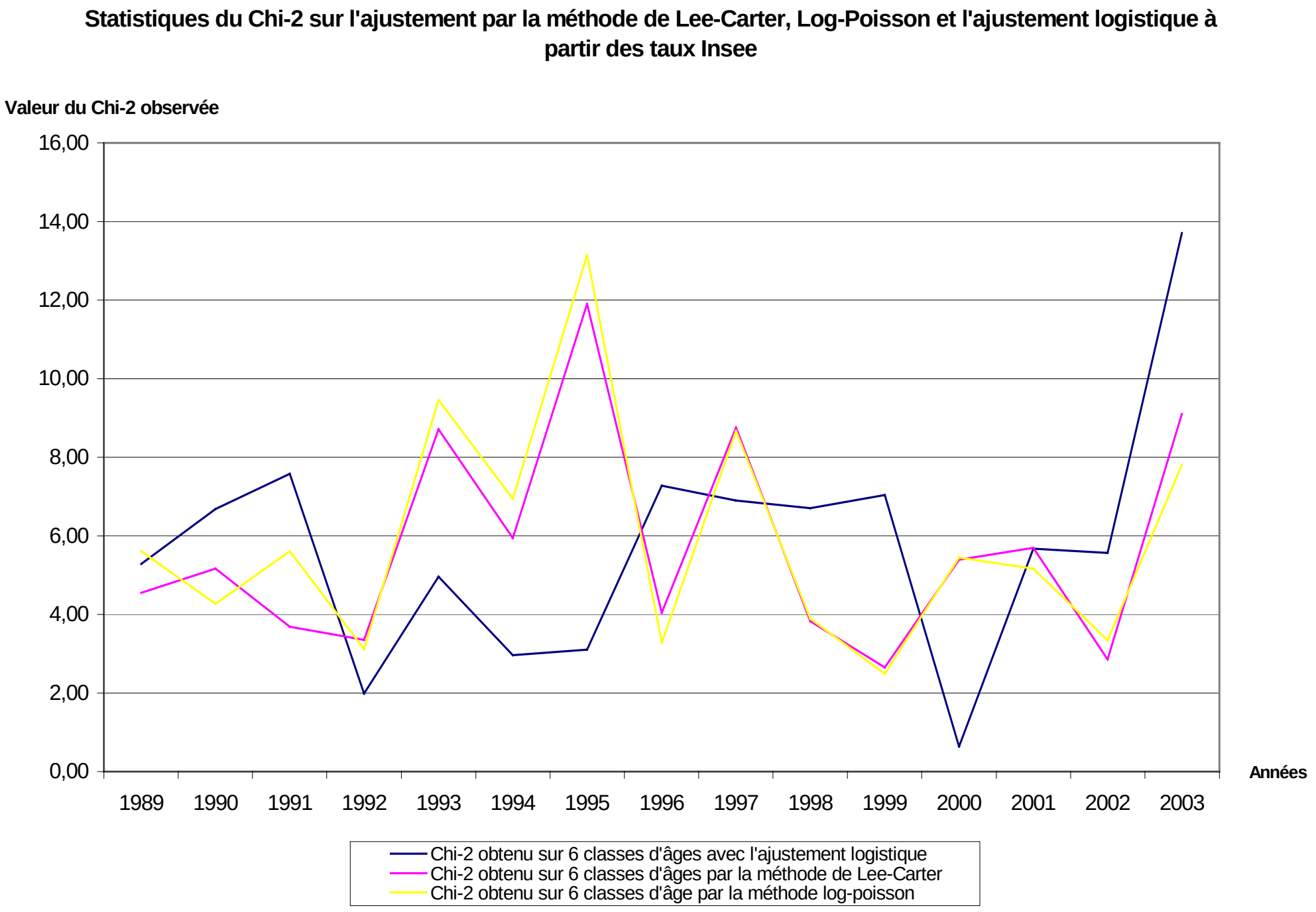

Fig. 12 : *Statistiques du Chi-2 entre les données brutes et l'ajustement logistique (à partir des taux Insee), par la méthode Lee-Carter, par la méthode log-poisson (avec l'ajustement polynomial des paramètres).*

On constate que la méthode de lissage par ajustement des taux logits à partir des taux logistiques INSEE produit en moyenne des taux légèrement plus proches des taux bruts que les méthodes de Lee-Carter et log-Poisson. La courbe bleue, correspond au chi-2 observé entre les décès lissés par l'ajustement logistique et les décès bruts ; elle est en effet généralement sous les courbes jaune et rouge (lissages obtenus par les modèles log-Poisson et Lee-Carter respectivement).

L'approche logistique apparaît donc ici être la plus efficace, tant du point de vue de la régularité des taux ajustés, que de la facilité d'extrapolation ou l'adéquation aux taux bruts.

## 2.2. Conséquence sur les provisions

Il n'est pas ici le lieu de développer complètement cette analyse[8], mais il est utile de fournir quelques éléments qualitatifs. Le modèle log-Poisson (qui doit être préféré systématiquement à Lee-Carter comme on l'a vu) s'appuie pour projeter les taux de mortalité futurs sur une extrapolation de la dérive temporelle *k*. Que l'on reste sur une détermination non paramétrique de cette tendance ou que l'on contraigne le modèle, on trouve une évolution globalement linéaire. Cette poursuite dans le future d'une tendance passé aussi régulière, sans infléchissement peut apparaître irréaliste. Toutefois, dans une approche prudentielle du risque, elle conduit *a priori* à une vision majorée des provisions.

Cette remarque peut être illustrée en considérant les capitaux constitutifs sur une tête $a_x = \sum_{k=0}^{\infty} v^k \, {}_k p_x$ où ${}_k p_x$ désigne la probabilité pour un individu d'âge $x$ d'être en vie dans $k$ années (l'année du calcul étant fixée arbitrairement). On suppose que le taux d'actualisation est constant égal à 2,25 %. On trouve alors qu'en 2003, sur nos données, on a :

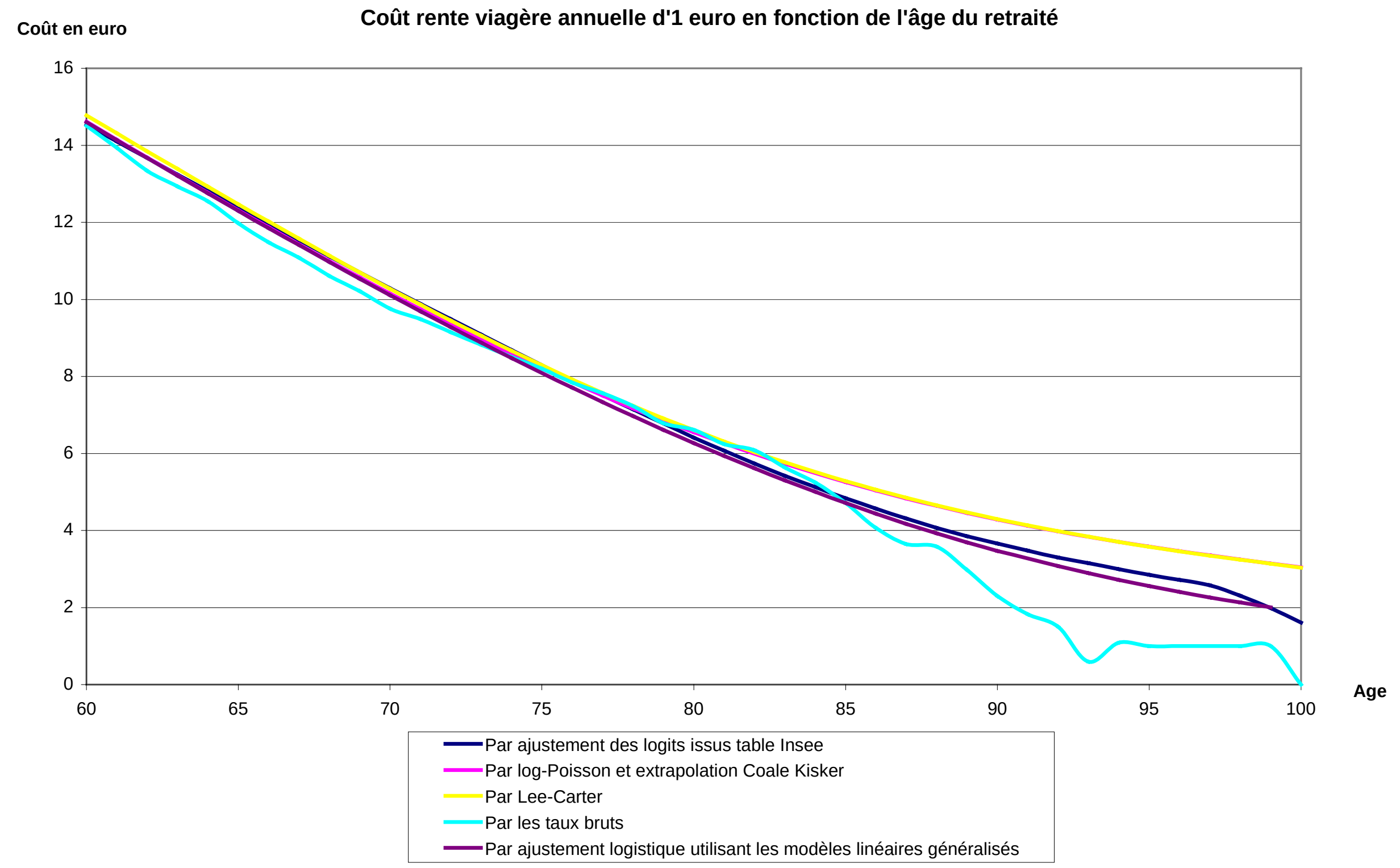

Fig. 13 : *Comparaison du coût des rentes viagères en 2003 suivant l'âge et la table employée.*

[8] Se reporter à LELIEUR [2005] pour une analyse plus complète.

Ce graphique illustre le caractère prudent, par construction, du modèle log-Poisson (et Lee-Carter), notamment par rapport à l'approche logistique. On y lit également le caractère inexploitable des données brutes.

Toutefois, il peut sembler préférable que les considérations prudentielles soient intégrées en aval de la réflexion, et ne viennent pas directement impacter le choix du modèle statistique. Pour cette raison, nous ne tirerons pas de conclusion de cette brève analyse.

# 3. Conclusion

La construction de tables de mortalité d'expérience permet de comparer la mortalité de la population assurée à celle de la population française en générale et de choisir la table la mieux adaptée pour tarifer et provisionner des rentes viagères.

A partir de données brutes, la création de tables de mortalité prospectives prend en compte les deux composantes, âge et année, pour l'estimation des taux de mortalité, et permet la projection de la mortalité future de la population. Deux méthodes différentes, Lee-Carter et log-Poisson, ont été appliquées dans ce document. Le modèle de Lee-Carter ne prend pas en compte l'hétéroscédasticité des résidus alors que sa variante log-Poisson permet de s'affranchir de l'hypothèse d'hétéroscédasticité et met au surplus en œuvre une estimation par maximum de vraisemblance[9].

La construction de ces tables prospectives présente la particularité dans cette étude de s'effectuer sur des portefeuilles de taille réduite. Cela implique des difficultés non rencontrées sur les précédents travaux, lorsque ces modèles étaient utilisés sur des populations ayant une échelle nationale.

D'une part, la volatilité des taux de mortalité bruts observés, liée aux fluctuations d'échantillonnage, entraîne une instabilité des trois coefficients permettant le calcul des taux de mortalité par les méthodes prospectives. On est ainsi conduit à diminuer le nombre de degrés de liberté de ces paramètres. Le fait d'imposer une structure polynomiale aux paramètres a finalement permis un meilleur lissage de la mortalité observée, en réduisant de manière drastique le nombre de paramètres du modèle.

D'autre part, aux âges les plus élevés, au-delà de 90 ans, on ne dispose plus de données brutes sur une population de taille suffisante. Cette caractéristique explique l'intervention de la méthode d'extrapolation de Coale-Kisker pour obtenir les taux de mortalité aux âges avancés.

Les résultats obtenus par l'utilisation de ces deux modèles montrent qu'en présence de l'ajustement polynomial des paramètres et lorsqu'on applique les méthodes sur les effectifs bruts (sans normalisation à 100 000 personnes à la naissance), les taux de mortalité obtenus sont très proches.

---

[9] A condition toutefois de ne pas renormaliser les effectifs et de travailler directement sur les effectifs exposés au risque.

Le modèle log-Poisson, associé à l'extrapolation de Coale-Kisker, apparaît meilleur sur la période 1989-2003 (période d'étude) que le modèle de Lee-Carter. Au surplus, il est plus prudent que celui de Lee-Carter pour l'estimation de rentes viagères sur cette période.

Afin de remédier aux inconvénients associés à la petite taille de l'échantillon, il est également possible de se tourner vers des modèles à référence externe. Le modèle retenu consiste à ajuster les taux logistiques bruts linéairement en fonction des taux logistiques des tables INSEE. L'ajustement est réalisé de telle sorte que les espérances de vie à 60 ans par les taux bruts et lissés soient équivalentes. Ce modèle permet d'obtenir un lissage conduisant par construction à des montants de provisions très proches de ceux issus des données brutes, mais présente parfois l'inconvénient d'être moins prudent que les modèles de Lee-Carter et log-poisson pour l'estimation du coût des rentes viagères.

D'autres techniques de lissage que celles présentées ici auraient pu être appliquées, on pense notamment aux ajustements par splines (DE BORR [1978]) ainsi qu'à la méthode non paramétrique de Whittaker-Henderson (TAYLOR [1992]), qui présente l'intérêt d'appartenir à la famille des lissage bayésiens.

Ces modèles ont fait l'objet d'une étude dans le cas de l'arrêt de l'arrêt de travail (WINTER et PLANCHET [2005]). Dans ce cas précis, l'arrêt de travail, les lissages obtenus ainsi sont de meilleure qualité que ceux issus des modèles Lee-Carter et log-Poisson (la qualité du lissage étant déterminée par la proximité des espérances mathématiques entre la table brute et la table lissée).

Enfin, on soulignera que l'ensemble des modèles discutés ici se prête aisément à une approche stochastique, pour la mesure du risque systématique de mortalité auquel se trouve confronté un régime de rentiers ; des travaux sont en cours afin de proposer un modèle simple permettant de quantifier ce risque, et notamment d'en comparer l'importance par rapport au risque financier auquel est également soumis le régime.

Ils fournissent une alternative intéressante aux modèles à structure affine d'origine financière souvent proposés pour la modélisation de ce risque.

# Bibliographie